\documentclass{article}
\usepackage{amsmath,amssymb}
\usepackage[dvips]{graphicx}
\begin{document}

\pagestyle{plain} 
\setcounter{page}{1}
\setlength{\textheight}{700pt}
\setlength{\topmargin}{-40pt}
\setlength{\headheight}{0pt}
\setlength{\marginparwidth}{-10pt}
\setlength{\textwidth}{20cm}

\title{Comments  on  Six Degrees of Separation based on the le Pool and Kochen Modelsgendary }
\author{Norihito Toyota \and Hokkaido Information University, Ebetsu, Nisinopporo 59-2, Japan \and email :toyota@do-johodai.ac.jp }
\date{}
\maketitle
\begin{abstract}
In this article we discuss six degrees of separation, which has been suggested by Milgram's famous experiment\cite{Milg},\cite{Milg2}, from a theoretical point of view again. 
Though Milgram's  experiment was partly  inspired to  Pool and Kochen's study \cite{Pool} that was made 
from a theoretical point of view. 
 At the time  numerically detailed study could not be made 
because computers and important concepts, such as the clustering coefficient,  needed for a network analysis nowadays, have  not yet developed. 
 In this article we devote deep study to the six degrees of separation based on some models proposed by  Pool and Kochen  by using a computer, numerically. 
 Moreover we  estimate the clustering coefficient along the method  developed by us \cite{Toyota1}
and extend our analysis of the subject through marrying Pool and Kochen's models to our method.

 \end{abstract}
\begin{flushleft}
\textbf{keywords:}
Six Degrees of separation,  Small world Network, Propagation Coefficient Model, Clustering Coefficient, 
\end{flushleft}

\section{Introduction}
\hspace{5mm} In 1967, Milgram made a great impact on  the world by advocating the concept 
 "six degrees of separation" by  a celebrated paper \cite{Milg} written based on an social experiment. 
"Six degrees of separation"  shows that people have a narrow circle of acquaintances. 
A series of social experiments made by him and his joint researcher\cite{Milg2} suggest 
that all people in USA are connected through about 6 intermediate acquaintances.   
Their studies were strongly inspired by  Pool and Kochen's study \cite{Pool}.    
At the time, however,  numerically detailed study \cite{Pool} could not be made 
because computers and important concepts, such as the clustering coefficient,  needed for a network analysis nowadays, have  not yet developed sufficiently.

One of the most refined models of six degrees of separation was formulated in  work of Watts and Strogatz\cite{Watt1},\cite{Watt2}.  Their framework provided compelling evidence that the small-world phenomenon is pervasive in a range of networks arising in nature and technology, and a fundamental ingredient in the evolution of the World Wide Web. 
 But  they do not examine closely  Milgram's original findings by their model:, especially how influence can 
the clustering coefficient proposed in their paper \cite{Watt1} have.  
We have made a study of them in our previous paper \cite{Toyota1} based on  a homogeneous hypothesis on networks. 
As a result, we found that the clustering coefficient has not any decisive effect on the propagation 
of information on a network and then information easily spread to a lot of people even in the cases with 
a relatively large clustering coefficient; a person only needs dozens of friends.  

 In this article we devote deep study to the six degrees of separation based on some models proposed 
by  Pool and Kochen \cite{Pool} by using a computer, numerically. 
 Moreover we  estimate the clustering coefficient along the method  developed by us \cite{Toyota1} 
and extend our analysis of the subject through marrying Pool and Kochen's models to our method. 
As a result, it seems to be difficult that six degrees of separation is realized in the models proposed by Pool and Kochen\cite{Pool} on the whole. 

The plan of this article is as follows. 
In the next section we argue on the first idea proposed by  Pool and Kochen\cite{Pool},   
where they impose a hypothesis on the average number $m_j$ of acquaintances common to $j$ individuals. 
In the section we give the condition that information spreads to  about $10^9$ people 
within  the parameters introduced in the model  and  evaluate how many 
 people can  receive  information released from one person after $6\sim10$ steps 
of transmission of information. 
We also argue that the results given by Pool and Kochen are unstable. 
In section 3, we study their improved model. 
We there investigate the almost same subjects as ones in the section 2.  
 In section 4, we apply the propagation coefficient model proposed by us \cite{Toyota1} 
 to Pool and Kochen\cite{Pool} to evaluate the clustering coefficient. 
 Then we discuss whether  six degrees of separation is feasible in Pool and Kochen's models\cite{Pool}. 
 The section 5 is devoted to summary and consideration.

\section{Pool and Kochen Model}
\hspace{5mm} 
Some models for patterns of social contacts were described in Pool and Kochen's paper \cite{Pool}. 
They can be broadly classified into two groups: 
the model with social strata and the ones without social  strata in the population considered. 
We  concentrate our discussion on  the former cases, which are described in the section 
 "The number of common acquaintances" of their paper. 
There mainly two considerations except for trivial notions are, mainly, described.  

$N$ is the total population in the region considered.   
A unique characteristics in their model is to introduce the average number $m_j$ of acquaintances   common to $j$ individuals such as illustrated in Fig.1. 
They assumed the following relation between $m_{ j+1} $ and $ m_j$; 
\begin{equation}
m_{ j+1} = a m_j \;\;for\; j=1,2,3,4,\cdots \;\;with\;\; 0<a<1.  
\end{equation} 
This means that for example the average number of acquaintances common to five people is smaller than the average number common to four by a factor $a$, which is the same proportion as the number of friends shared by four is to the number shared by three. 
This $a$ is between 0 and 1 and should be statistically estimated. 
It is also assumed that $a$ is independent of  $j$
\begin{center}
\includegraphics[scale=0.8,clip]{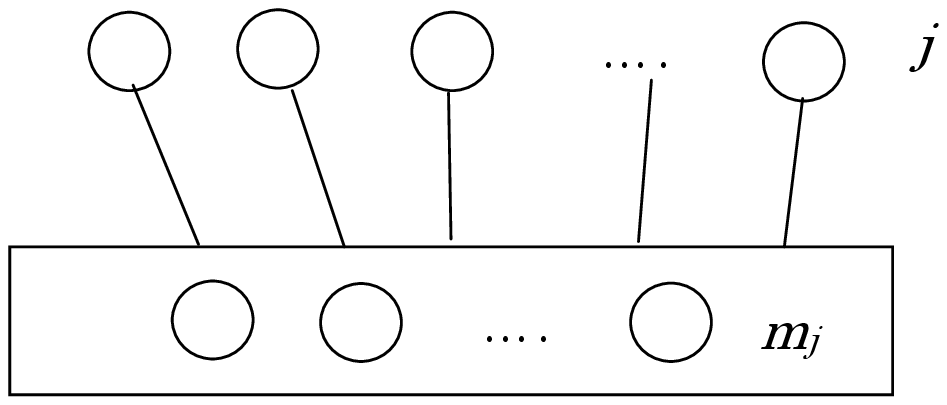} \\
Fig．1. A schematic illustration of $m_j$.　　
\end{center}　

When 
$n_i$ is the number of independent acquaintances standing  at intervals 
of $i$ link length away from a person A ($i$-$th$ generation from A), we can express $P_i$, 
which is the probability that  acquaintances  standing at $i$-$th$ generation from the person A 
are just an  acquaintance of another person B chosen randomly from people 
except for the acquaintance tree starting from A , 
as 
\begin{equation}
P_i= \frac{n}{Na} \prod_{j=0}^{i-1} ( 1-P_j) \left\{ 1-(1-a)^{n_{i}}  \right\} . 
\end{equation}
The Fig.2 shows the situation of Pool and Kochen Model. 
The following recursion relation holds for $n_i$,
 \begin{equation}
n_{i+1}=\frac{n}{a}  \left\{ 1-(1-a)^{n_{i}}  \right\}, 
\end{equation}
where $n$ is the average number of acquaintances that any person knows 
(referred as the propagation coefficient in our paper\cite{Toyota1})  
and  satisfies $m_1=n$. 
By solving this recursion relation, we can  find the total population $M(d)$ that receive 
information propagated from A  during $d$ generations;
\begin{equation} 
M(d)= \sum_{k=1}^{d} n_k . \nonumber
\end{equation}

Paying attention  to $m_1=n_1=n$, we obtain 
\begin{equation}
n=\frac{Na^2}{ 1-(1-a)^{n} }, 
\end{equation}
which leads to a useful relation $a$ and the propagation coefficient $n$.  

\begin{center}
\includegraphics[scale=0.8,clip]{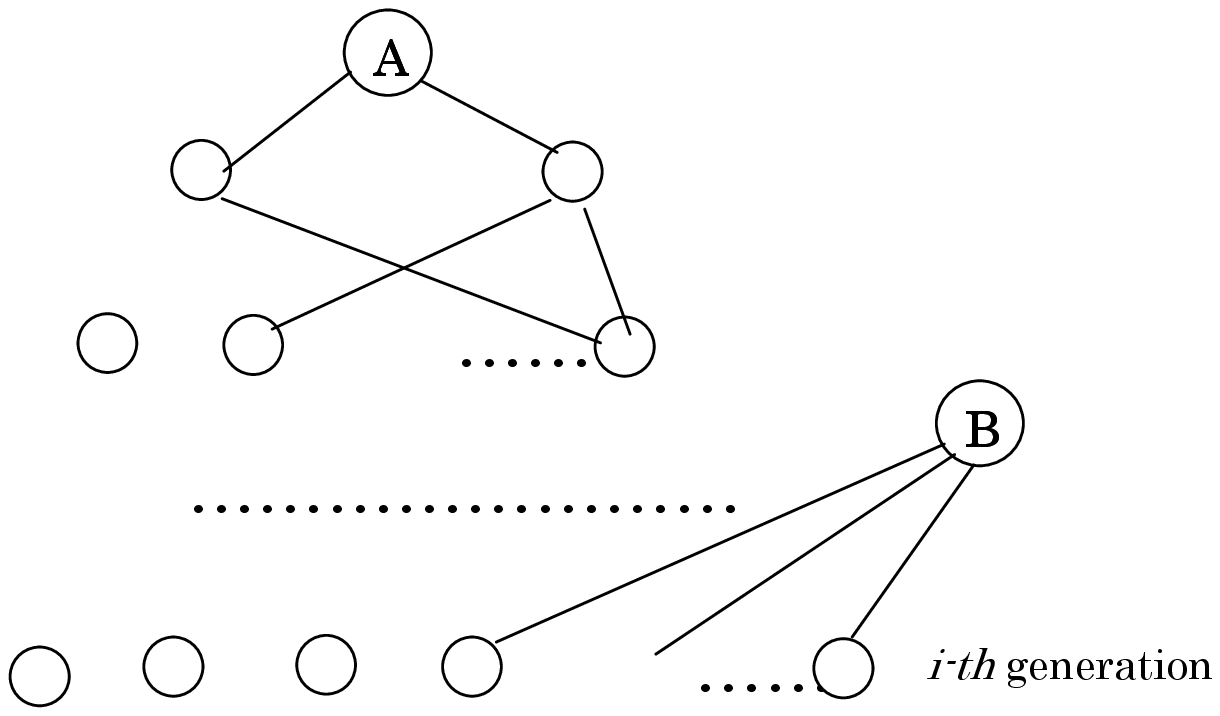} \\
Fig．2. A schematic illustration  of Pool and Kochen Model.　　
\end{center}　

Fig.3 shows the numerical relation of Eq.  (4) for $N=10^9$. 
The value of $N$ comes from a slightly large population of the USA. 
From this, in order to obtain a realistic value of $n$ for $N=10^9$, 
 extremely small values of $a$ are  needed  
such as $a\sim O(10^{-6})$  when $n\sim O(10)$, and $a\sim O(10^{-4})$ when $n\sim O(10^{2\sim 3})$. 
We refer to Bernard et al. \cite{Bernard1,Bernard2,Bernard3}, 
where they estimate that the average  person has a social circle of about 290 people from empirical studies. 
We infer $a\sim $ (a few)  $\times 10^{-4}$ from that value.

From these results, evaluating the total propagation population $M$ 
under appropriate values of $a$ and $n$, 
we obtain the following results;
$M \doteqdot 4\times 10^5$ for $a\sim 10^{-7}$ and $n=10$, and 
  $M\doteqdot 6\times 10^6$ for $a=10^{-3}$ and $n=1346$. 
 This means that $M$ evaluated is insufficient for information even to spread to 
only one percent of the total population  even in the case with $M\sim 6\times 10^6$.    
Since $n_i$ rapidly converges to a constant when generation $i$ grows into about $3$ in the case of $n=1349$, such as shown in Fig. 4,
   $M$ grows larger in proportion to generation number $d$ for $d>3$. 
 This is a reason that information can not spread over a large population.

\begin{center}
\includegraphics[scale=0.6,clip]{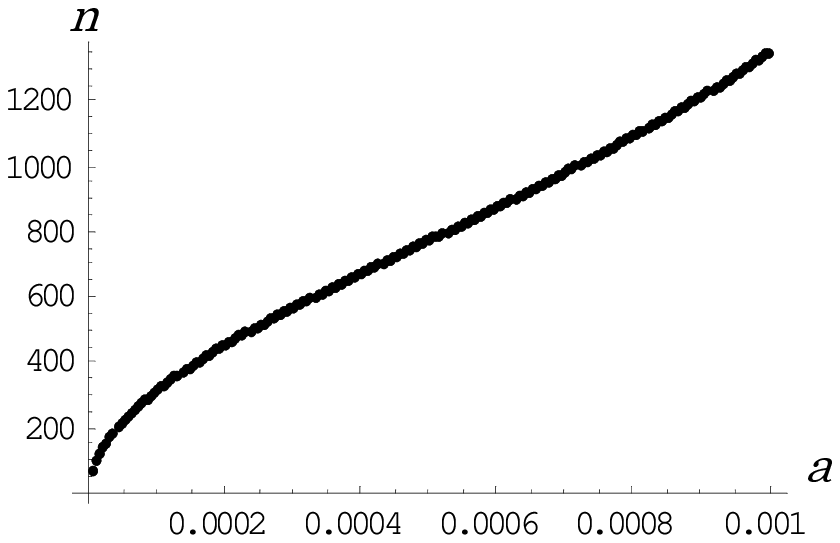} \hspace{15mm} \includegraphics[scale=0.6,clip]{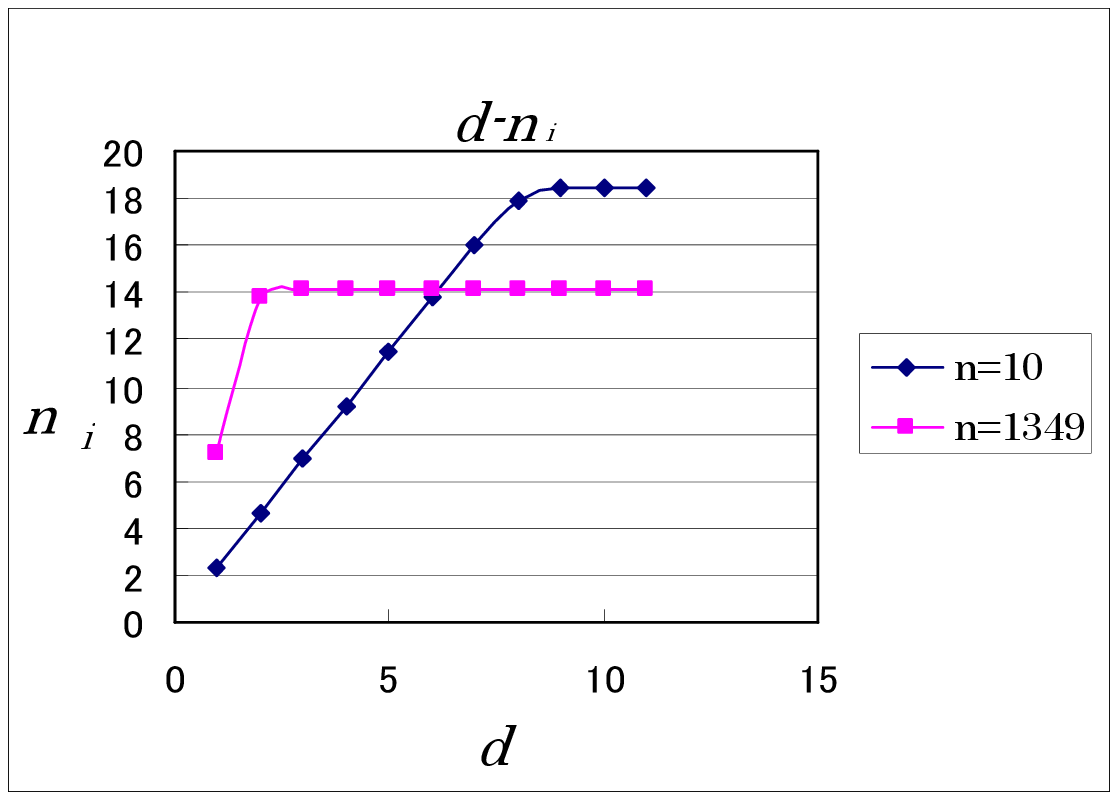} \\
Fig.3 $a$-$n$ plot for  $N=10^9$.　　 \hspace{15mm} Fig.4 $d$-$n_i $ plot for  $a=10^{-7}$,  $n=10$ and  \hspace*{50mm} $a=10^{-3}$,  $n=1346$.
\end{center}

\begin{center}
\includegraphics[scale=0.8,clip]{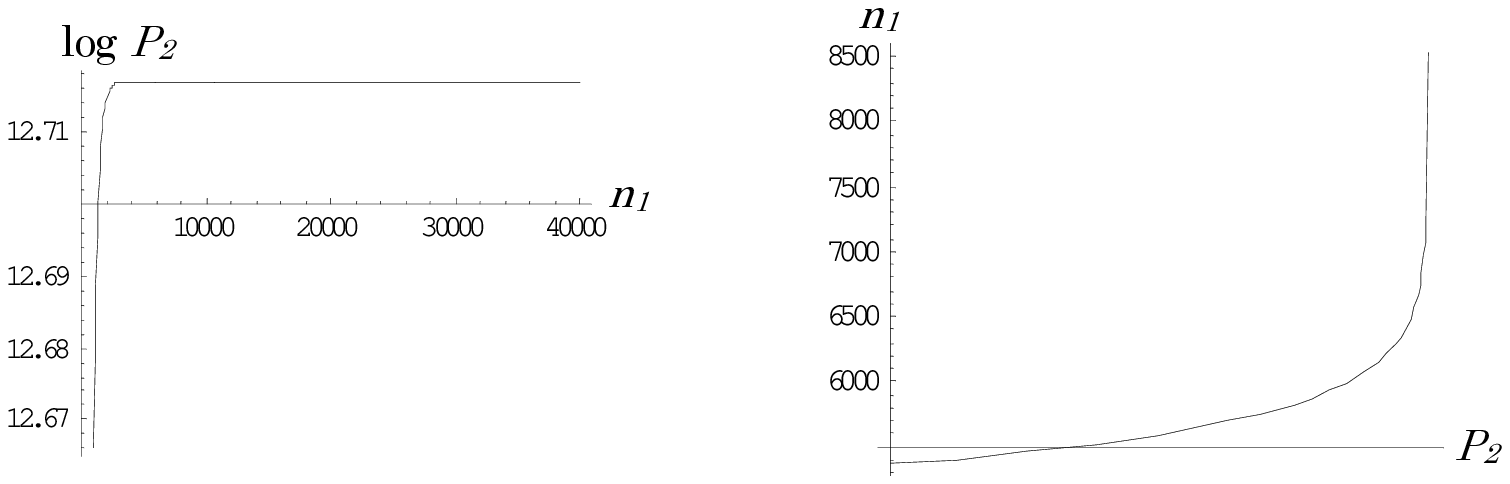} \\
Fig.5. \hspace{5mm} fig.5a.  $n_1$-$\log P_2$  plot  \hspace{25mm}  fig.5b $\log P_2$-$n_1$ plot with the interval 
of \\ \hspace{90mm}$P_2 =[0.003333333,0.003333334]$.　　
\end{center}　
Poor and Kochen mainly gave  analyses of the relation between  $P_i $ and $n_i$. 
Analyzing numerically it in some detail, we find that the relation so unstable that it is difficult 
to get reliable claims. 
As shown in Fig.5, $n_1$ rapidly grows  around $P_2=1/3$ and that is to say, $P_2$ rapidly grows 
for small changing  of $n_1$. 
Though the relation between  $P_i $ and $n_i$ is unreliable around there, the relation between  $a $ and $n$ is stable as shown in Fig.3.  
Thus the estimation of $M$ calculated above is also reliable.

\section{Version up Model}
\hspace{5mm} As the second step, Poor and Kochen developed their model more fully. 
There they introduce a set $K_A$ of A's circle of acquaintances and its complement$\bar{K}_A$. 
$A_i$ denote the individuals in the set $K_A$.   
The following assumption  are made on the conditional probability $Prob(B \in \bar{K}_{A_k} | B \in \bar{K}_{A_{k-1}},B \in \bar{K}_{A_{k-2}},\cdots, B \in \bar{K}_{A_1} )$; 
\begin{equation}
Prob(B \in \bar{K}_{A_k} | B \in \bar{K}_{A_{k-1}},B \in \bar{K}_{A_{k-2}},\cdots, B \in \bar{K}_{A_1} )=
Prob ( \bar{K}_{A_k} |   \bar{K}_{A_{k-1}} )=b=const.  
\end{equation}
where B is a person randomly chosen and the constant $b$ should be statistically estimated. 
Thus we get \cite{Pool} 
\begin{equation}
Prob(\bar{K}_{A_k}, \bar{K}_{A_{k-1}}, \cdots, \bar{K}_{A_1} )=Prob(\bar{K}_{A_1}) b^{k-1}=(1-\frac{n}{N})b^{k-1}. 
\end{equation} 
Since for  $k=2$   
\begin{equation}
Prob(\bar{K}_{A_2}, \bar{K}_{A_{1}} )=(1-\frac{n}{N} )b=1-\frac{2n}{N}+\frac{m_2}{N},
\end{equation} 
so we have 
\begin{equation}
b= \frac{1-\frac{2n}{N}+\frac{m_2}{N}}{1-\frac{n}{N} }.
\end{equation}

From these equations, we get  \cite{Pool} 
\begin{eqnarray}
n_{k+1}= \frac{n^2}{m_2}  \left\{  1-(1-\frac{m_2}{n})^{n_k}  \right\}, \\
P_k=  \prod_{i=0}^{k-1} \left\{ 1-P_i \right\}P_k^\prime,\\
P_k^\prime = 1-(\frac{n}{N} ) b^{n_{k-1}} .
\end{eqnarray}

By using these relations, Poor and Kochen mainly studied about $P_k$, but 
did not give no consideration to $M$.  
In order to do it we only need to solve the recursion relation (9). 
In this article we numerically estimate $M$ for $N=10^9$, changing values of $n$ and $m_2$,. 
The results are partly given by Fig.6. 
 As expected, $M$ increases as $n$ becomes larger. 
We also find a natural result that the more larger  $m_2$ is, the smaller $M$ is.  
It is, however, impossible that information does  spread to  most of the total population even after 10-th generation in the both cases of Fig.6. 
\begin{center}
\includegraphics[scale=0.8,clip]{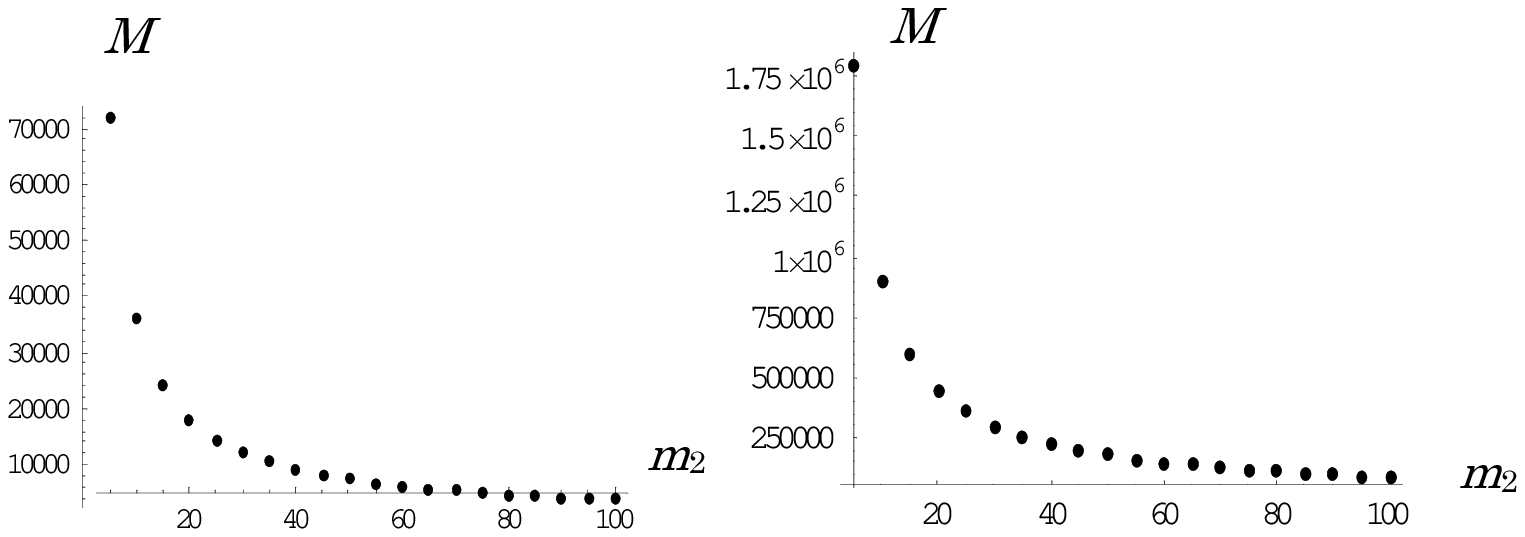} \\
Fig.6. Total population $M$ for $n=200$  (left) and $n=1000$  generation (right) in $10$-$th$. 
\end{center}　

\section{Poor-Kochen Model and Clustering Coefficient}
\hspace{5mm} 
In this section we calculate the clustering coefficients in Poor-Kochen Model 
according to the general method developed in the propagation model \cite{Toyota1}. 
First of all, we tidy the notations used in the model. \\
$ n_i  $ is the number of nodes in $ i$-$th $ generation $G_i$.\\
$N$ is the number of total nodes or the size of a network.\\
$\overline{C}$ is the clustering coefficient of a network.\\
$C_i$ is the  contribution to the clustering coefficient produced  in $G_i$. \\
$ k_{i,i}$ is the number of edges  connected between the same generation $i$.  \\
$ k^{(j)}_{i,i+1}$ the number of edges  from a node $j$ in $G_i$  to nodes of  $ G_{i+1} $.\\
$\overline{k}_{i,i+1}$ is the average of $ k^{(j)}_{i,i+1}$ over all nodes relevant to the generation;
\begin{equation}
\overline{k}_{i,i+1} = \frac{\sum_{j\in G_i}^{n_i} k^{(j)}_{i,i+1}}{n_i}.
\end{equation}

Using these quantities, we can express the average degree $K$ of a network. 
\begin{equation}
K  = 1+\frac{2k_{i,i}}{n_i} +  \overline{k}_{i,i+1}  
\end{equation}
In the propagation model we assumed that $C_i$ was constant, but we now should assume that 
the population $n_i$ in each generation is constant in Pool-Kochen model. 
So assuming that the recursion relation on  $n_i $ satisfies  
\begin{eqnarray}
n_{i+1} &=&   \overline{k}_{i,i+1}  n_i  (1-q)=const.\equiv \bar{n},
\end{eqnarray}
we have 
\begin{equation}
 \overline{k}_{i,i+1} =\frac{1}{1-q},
\end{equation}
where $q$ denotes the probability that a node has two acquaintances in the generation earlier 
than  the node.  
A schematic diagram of the propagation model including the parameter $q$ is given by Fig.7.  
Substituting  these expressions into the  following equation (16) given by us\cite{Toyota1}
\begin{equation}
C_i=\frac{ \overline{k}_{i-1,i} n_{i-1}  }{K(K-1)} 
\biggl( \frac{( \overline{k}_{i-1,i} -1)(K-1-\overline{k}_{i,i+1})}{ \overline{k}_{i-1,i} n_{i-1} -1 } +
\frac{2q(K-1-\overline{k}_{i-1,i} )} {n_{i-1}}\biggr),
\end{equation}
We obtain 
\begin{equation}
C_i (q)=\frac{ \bar{n} q  }{(1-q)^2 K(K-1)(\bar{n} -1-q)(\bar{n} -1)}    \biggl( (\bar{n} -1)(K-q K-1)+2(K-q K+q-2)(\bar{n} -1+q)\biggr).
\end{equation}

When varying $q$, the behavior of $C_i$ in the propagation model is shown in Fig. 8 
where $K=200$ and $\bar{n}=1000$ are taken.  
This value will be proper for inhomogeneous networks with respect to degree distribution 
, which is actually assumed  in this section. 
For larger values of $K$ and $n$ such as  $K=1000$ and $n=150000$, 
we have checked that  $C_i$  increases more rapidly as $q$ grows larger. 
Thus the clustering coefficient does not grow large unless $q$  considerably becomes large.  
It is thought to be difficult that Pool-Kochen model can realize any small worlds  
 from the perspective of this analysis, too. 
So far networks based on Pool-Kochen models do  display a large world property 
and small clustering coefficient is preferable.   

\begin{center}
\includegraphics[scale=0.8,clip]{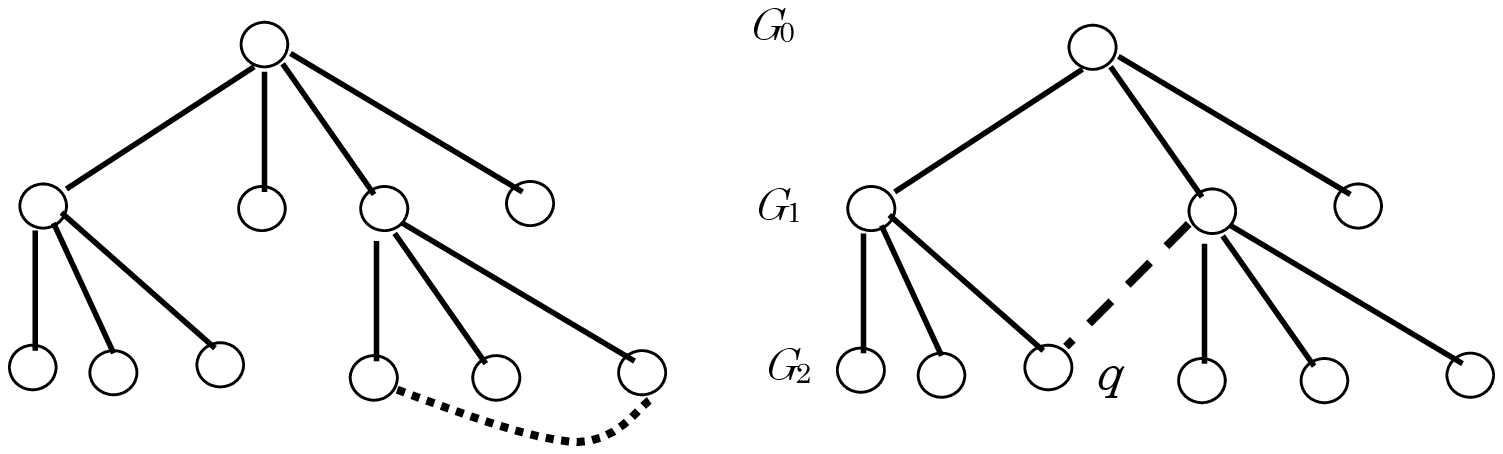} \\
Fig 7. A schematic diagram of the propagation model. 
\end{center}　

\begin{center}
\includegraphics[scale=0.8,clip]{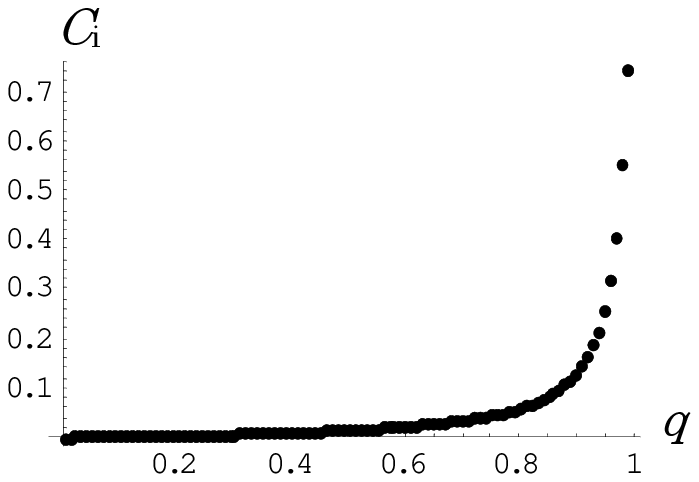} \\
Fig 8. $q$-$C_i$ plot for $K=200$ and $\bar{n}=1000$.  
\end{center}　

Pool and Kochen have made some discussions  on models with social strata. 
Since it is, however, though that the models do not bring any correct results, we  abandon the pursuit of the models.

\section{Summary and Consideration}
\hspace{5mm} 
In this article we numerically analyzed how six degrees of separation can be realized in human networks  based on a series of Pool-Kochen models. 
Moreover we estimate the clustering coefficient of Pool-Kochen models according to the propagation model and explored  the possibility of small-worldness. 

In result, we found that it is difficult that Pool-Kochen models realize  six degrees of separation and 
also achieve a large clustering coefficient. 
Recently Kleinfield has fanned some critical discussions to Milgram's empirical evidence for six degrees of separation \cite{Klein}. 
Later Watts ea al. have conducted  by far the largest ever small-world  experiment by using E-mail, 
involving  60 thousand E-mail users with targets over 13 countries  \cite{Watt3,Watt4}. 
 As they recognize,   their experiment has a positive bias in the choice of  E-mail users. 
The small world problem, however, remains as fascinating psychological mysteries. 
In my opinion, the meaning of Pool-Kochen's study is that 
the models  will not do much  a better understanding of six degrees of separation but 
inspired Milgram and so on to study  interesting subjects such as six degrees of separation. 

Once we  discussed it based on homogeneous hypothesis in the propagation model \cite{Toyota1} 
and find that six degrees of separation likely to materialize somewhat. 
The hypothesis is, however, no correct. 
There is considerably deviation  in the degree distribution, the clustering coefficient and so on in real networks.   
 To understand  six degrees of separation more really, we should  introduce some correct distributions into 
  the degree distribution, the clustering coefficient and so on. 
  Newman \cite{Newm21} has made discussions on it by considering some distribution in the degree distribution and 
  furthermore "mutuality" which is a quantity that reflects the density of squares in human relations.  
This direction of study seems to play an important role in the studies of  six degrees of separation. 
The detail research  toward the line of this  should be made more properly.

\end{document}